\documentclass[preprint,12pt]{article}
\usepackage{subcaption}
\usepackage{svg}
\usepackage{amssymb}
\usepackage{arxiv}
\usepackage[T1]{fontenc}
\usepackage{hyperref}
\usepackage{url}
\usepackage{booktabs}
\usepackage{nicefrac}
\usepackage{microtype}
\usepackage{doi}
\usepackage{amsfonts}
\usepackage{graphicx}
\usepackage{epstopdf}
\usepackage{algorithmic}
\usepackage{mathtools}
\usepackage{bm}
\usepackage{tikz}
\usetikzlibrary{arrows,positioning}
\usepackage{pgfplots}
\usepackage{amsopn}
\usepackage{amsmath}
\DeclareMathOperator*{\argmin}{arg\,min}
\usepackage[scientific-notation=true]{siunitx}

\title{Globally Adaptive and Locally Regular Point Discretization of Implicit Surface Geometries}

\date{}

\author{ \href{https://orcid.org/0000-0003-2915-8920}{\includegraphics[scale=0.06]{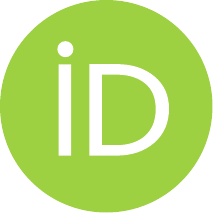}\hspace{1mm}Lennart J.~Schulze}\\
	\texttt{lschulze@mpi-cbg.de}
	\And
	\href{https://orcid.org/0000-0003-4414-4340}{\includegraphics[scale=0.06]{orcid.pdf}\hspace{1mm}Ivo F.~Sbalzarini}\thanks{Corresponding author; now at: Department of Mathematical Modeling and Machine Learning, University of Zurich, Zurich, Switzerland.}\\
	\texttt{ivo.sbalzarini@uzh.ch}}

\renewcommand{\headeright}{}
\renewcommand{\undertitle}{}
\renewcommand{\shorttitle}{Adaptive Point Discretization of Implicit Surfaces}

\hypersetup{
pdftitle={Adaptive Point Discretization of Implicit Surfaces},
pdfauthor={L.J. Schulze and I.F. Sbalzarini},
}

\begin{document}
\maketitle
\setcounter{footnote}{0}
\vspace{-1cm}
\begin{center}
	Dresden University of Technology, Faculty of Computer Science, Dresden, Germany,\\Max Planck Institute of Molecular Cell Biology and Genetics, Dresden, Germany,\\Center for Systems Biology Dresden, Dresden, Germany
\end{center}
\vspace{2cm}
\begin{abstract}
Point discretization of implicitly defined curved surfaces is required throughout computer-aided design, engineering analysis, and scientific computing. These applications often impose that  surfaces are sampled with local regularity and global curvature adaptivity for unambiguous geometry representation. Finding a numerically well-conditioned point discretization, however, is not trivial, even for simple analytic surfaces. We present an algorithm for finding near-optimal surface point distributions governed by a prescribed resolution field on a curved surface. The algorithm works by approximately minimizing a global potential over local point--point interactions. We leverage a high-order meshfree level-set method to implicitly describe the surface and to perform all required projections without additional surface-attractive forces. To accelerate convergence, the algorithm dynamically fuses and inserts points where a local excess or lack is detected, for which it introduces an integral support measure. We test the proposed algorithm on a variety of shapes, ranging from parametric to non-parametric surfaces. We find point discretizations with different curvature adaptivity and low deviation from the prescribed target spacing. In all cases, the presented algorithm rapidly and robustly converges to the final number and distribution of surface points.
\end{abstract}


\section{Introduction}
Implicitly defined curved surfaces are ubiquitous in many engineering applications. Examples include level sets obtained from segmentation of micro-CT scans of material microstructures \cite{Petrasch2008b}, the construction of additive-manufacturing lattices such as triply periodic minimal surface (TPMS) gyroids \cite{al2019multifunctional}, and simulations involving fluid-structure interactions \cite{cottet2006level}. Many downstream tasks, such as mesh generation \cite{boubekeur2009mesh,alliez2008recent,Friess2013,khan2020surface}, surface quadrature \cite{venn2026high}, the solution of partial differential equations (PDEs) \cite{dziuk2013finite,singh2023meshfree}, and object rendering \cite{kobbelt2004survey} benefit from point discretizations of high local regularity. 

Placing points regularly, and ideally curvature-adaptively, on a given implicit surface is not only a fundamental problem in mesh generation, but in computational engineering in general. Most methods for interpolation, integration, and the approximation of derivatives are based on adequately placed sample points, nodes, or particles that discretize the domain of interest. In curved surfaces, intrinsic curvature introduces an additional requirement on the local resolution. Higher curvature requires locally finer resolution to well-sample the surface. Additionally, methods that use the embedding space, such as Surface Discretization-Corrected Particle Strength Exchange (Surface DC-PSE) \cite{schrader2010discretization,singh2023meshfree} require non-intersecting tubular neighborhoods, which also depend on local curvature. If the surface dynamically deforms, the size of the discretized domain can vary, introducing another challenge: Local expansion and contraction of the surface area requires adding and removing sample particles in order to maintain sampling quality. Point distributions that violate the requirements of downstream methods can lead to unphysical results, unstable computations, or render the problem entirely unsolvable. 

Owing to its fundamental importance, the problem of well-sampling curved spaces has been considered in several disciplines. In surface finite element methods \cite{dziuk2013finite}, for example, extensive research has been done for surface remeshing \cite{alliez2008recent}, where low-quality meshes, typically obtained from 3D scanning software, are improved to enhance triangle quality and  vertex regularity. The quality of the resulting unstructured mesh heavily depends on the regularity of the node distribution and the resulting volumes or surface areas of the elements \cite{nguyen2009constrained}. Some methods aim to improve the node distribution by locally removing extreme angles and edge lengths in the mesh \cite{dunyach2013adaptive}. Another approach to remeshing relies on mesh simplification \cite{heckbert1997survey}, which aims to reduce detail where redundant to optimize downstream computation and memory usage. A popular class of remeshing methods is based on the closely related Delaunay triangulation and centroidal Voronoi tessellation \cite{du1999centroidal}. Most surface-remeshing methods rely on a given mesh as input, but some of them have been extended to other settings, such as centroidal Voronoi tessellation of 3D point clouds \cite{chen2018point}. Triangulated surface representations have both advantages and disadvantages. Their explicit nature allows for exact computation of geodesic distances \cite{webb2025curvedspacesim,mitchell1987discrete} and guarantees that the node points (by definition) lie within the surface. However, triangulations sacrifice the smoothness of curved surfaces, such that computing normals and curvature by discrete differential geometry becomes harder \cite{duclut2022coarse}.

In addition to explicit surface representations, such as triangulations, the problem of sampling and meshing has also been extensively studied for implicitly defined surfaces \cite{de2015survey}. A classic approach is the physically-based particle system \cite{de1992physically}, which is related to the algorithm we propose here. It yields regular particle distributions on curved surfaces by computing equilibrium states of a system with attractive and repulsive forces. The attractive forces bind the particles to the zero level-set of the implicit surface, whereas the repulsive forces act between particles to distribute them homogeneously. After the algorithm has converged, a polygonization of the particles is performed to obtain a surface mesh, for example using the ball-pivoting algorithm \cite{bernardini2002ball}. This idea was further developed in the Witkin--Heckbert system \cite{witkin1994using}, which additionally included a ``birth and death'' routine after each relaxation toward equilibrium to accelerate overall convergence. Further, it added the possibility of manipulating the surface, making it an attractive method for computer graphics. The class of energy-based algorithms has since been extended to curvature-adaptive discretization \cite{crossno1997isosurface}, to surfaces with singularities \cite{rosch1997interactive}, and to incorporating automatic and numerical differentiation of the implicit function defining the surface \cite{hart2005using}. The approach has also been extended to second order using forces that depend on the Hessian of the particle--particle interactions \cite{meyer2005robust}. A related method for noisy data, where sample points cannot be guaranteed to lie within the surface, uses a moving least-squares representation of the surface \cite{alexa2003computing,pauly2002efficient}. 
A similar force-balance relaxation approach has also been proposed for volumetric domains \cite{zhu2021cad} and for solving PDEs using Lagrangian particle methods in the volume \cite{reboux2012self}. In the latter, the authors suggest minimizing the global potential by gradient descent augmented with a line search to adaptively determine the optimal step size. 

Despite these advancements, energy-based surface sampling often faces two challenges: (1) The rules for particle addition and removal can be unclear, with inadequate choices preventing convergence of the algorithm and causing oscillations \cite{rosch1997interactive}. (2) The optimization within a curved surface requires point positions and vector fields to lie in the tangent space. A surface-normal penalty force does not, in general, ensure that particles remain within the surface, and it requires careful parameter tuning \cite{meyer2005robust}.

Here, we specifically address the two challenges of particle insertion and removal, and of ensuring that particles stay within the surface to a high order of accuracy. Our approach leverages the classic principle of energy-based particle relaxation and the minimization method proposed in Ref.~\cite{reboux2012self} to compute locally regular but globally curvature-adaptive particle distributions on implicitly defined surfaces. We enforce the particles to lie within the surface by projecting them onto their respective closest surface points. We use the same closest-point transform to also project any vector-valued field on the particles into the tangent space of the curved surface. We accurately and efficiently compute the closest-point transform using the particle closest-point (PCP) method \cite{schulze2024high}. For addition and removal of particles, we propose a novel approach based on an integral support measure that is computed locally. We demonstrate the robust convergence of the resulting algorithm for computing adaptive particle discretizations of a variety of parametric and non-parametric shapes over wide ranges of resolution and curvature-adaptivity. We call the presented algorithm the {\em self-adaptive implicit surface sampling} (SAISS) method. A scalable parallel implementation is available as part of the open-source scientific-computing library {\tt OpenFPM} \cite{incardona2019openfpm} under \url{https://git.mpi-cbg.de/mosaic/software/parallel-computing/openfpm/openfpm_numerics}.

\section{The SAISS Method}

We aim to find a locally regular, curvature-adaptive particle discretization of a given surface. Different metrics can be used to define the target density of the particles, such as the surface area a particle should occupy or the number of neighboring particles within a certain radius. Here, we use the minimum nearest-neighbor distance per particle as the main input parameter for our method.
In addition, the user specifies how strongly surface curvature shall affect the particle distribution. This results in two input parameters to the method: (1) the global target or reference spacing $h_0$, and (2) the curvature weight $\tau$. 

Once these parameters are given, the number of particles required to discretize a certain surface is defined by the surface area and the surface integral over the curvature magnitude. An analytical expression for the total surface area or the curvature of arbitrary surfaces is not available. This makes it impossible to {\it a priori} know the number of particles $n_p$ in the target discretization. Therefore, $n_p$ is automatically determined by the SAISS algorithm.
The SAISS algorithm therefore jointly solves two problems: (1) find a curvature-adaptive surface relaxation of the particle set; (2) find an appropriate size of the particle set. This defines a non-convex (due to (1)) and non-differentiable (due to (2)) optimization problem, which we approximately solve in alternating iterations using operator splitting.

\subsection{Relaxing the particle distribution}
The two input parameters $h_0$ and $\tau$ imply a parametrized definition of the local target spacing $\Tilde{h}$ at a location $\mathbf{x}$ within the surface as:
\begin{equation}
    \Tilde{h}(\mathbf{x})=G_{h_0,\tau}(\mathbf{x}).
\end{equation}
In principle, $G_{h_0,\tau}(\mathbf{x})$ can be any continuous function that describes the desired adaptivity. Here, we focus on curvature adaptivity, which requires the spatial dependence to be a function of surface curvature. Specifically, we use the {\em deviation from flatness} $\sqrt{\kappa_1^2+\kappa_2^2}$ with principal curvatures $\kappa_1,\kappa_2$ as a smooth measure of local curvature. It is non-negative and continuously differentiable for smooth surfaces. We then define the parametrized local target spacing as:
\begin{equation}\label{eq:htilde}
    \Tilde{h}(\mathbf{x})=G_{h_0,\tau}(\mathbf{x})=G_{h_0,\tau}(\kappa_1(\mathbf{x}),\kappa_2(\mathbf{x}))=\frac{h_0}{\sqrt{1+\tau\left|\sqrt{\kappa_1^2+\kappa_2^2} - \kappa_\text{ref}\right|}}\, ,
\end{equation}
where $\kappa_\text{ref}$ is the reference curvature at which the resolution $h_0$ is assumed (typically, $\kappa_\text{ref}=0$).

In areas where the curvature, and hence $\Tilde{h}$, changes quickly, there may be significant resolution gradients between particles. Many mesh-free particle methods approximate operators locally within finite neighborhoods of radius $r^*$ around each particle $i$ at location $\mathbf{x}_i$. To ensure sufficient neighborhood support for both high- and low-curvature particles, we assign each particle a characteristic length
\begin{equation}\label{eq:characteristicLength}
    h_i=h(\mathbf{x}_i)=\min_{d_{ij}\leq r^*}\Tilde{h}(\mathbf{x}_j)
\end{equation}
as the minimum $\Tilde{h}$ in its $r^*$-neighborhood. Here, $d_{ij}$ is the distance between particles $i$ and $j$. From the resulting target resolution field $\{h_i\}_{i=0}^{n_p-1}$, we define the potential energy
\begin{equation}\label{eq:totalpotentialenergy}
    V_\text{tot}(\mathbf{x}_0,\ldots,\mathbf{x}_{n_p-1})=\sum_i\sum_jV_{ij}(\mathbf{x}_i, \mathbf{x}_j) \, ,
\end{equation}
over local pairwise interactions $V_{ij}$. There are many possible choices for the local pairwise potential, such as the well-known Lennard--Jones potential \cite{jones1924determination}, or purely repulsive potentials, such as the \textit{s-energy} potentials \cite{saff1997distributing}. While attractive-repulsive potentials are advantageous in open domains, attractive forces in the tangential direction are not required in closed surfaces. We therefore use the purely repulsive potential \cite{reboux2012self} 
\begin{equation}
V(r)=
\begin{cases}
    0.8\cdot2.5^{1-6r}-2.5^{-6r}\quad&\text{if }r<r^*,\\
    0&\text{else}
\end{cases}
\label{eq:potential}
\end{equation}
and integrate it into the particle system by rescaling with the minimum characteristic length per particle pair $h_{ij}=\min(h_i,h_j)$, such that the particle--particle potential reads
\begin{equation}
    V_{ij}=h_{ij}^2V(d_{ij}/h_{ij})
\end{equation}
with $V(r)$ from Eq.~\ref{eq:potential}.
Minimizing the resulting total energy in Eq.~\eqref{eq:totalpotentialenergy} relaxes the particle arrangement to become locally homogeneous with a resolution of at least $h(\mathbf{x})$. The overall optimization problem can be stated as
\begin{equation}\label{eq:globaloptimizationproblem}
    \argmin_{\{\mathbf{x}_i\}}\sum_i\sum_jV_{ij}
\end{equation}
for $i,j\in\{0,\ldots,n_p-1\}$. This optimization problem is highly non-convex with a number of local minima that is estimated to grow exponentially with $n_p$ \cite{erber1995comment}, which renders exact solution infeasible. Luckily, though, the local minima are close to the global minimum in terms of their total energy \cite{saff1997distributing}. Therefore, near-optimal local solutions suffice for downstream applications that only require a certain local regularity of the particle distribution, around which small distortions are allowed. 

We compute approximate solutions to the minimization problem in Eq.~\eqref{eq:globaloptimizationproblem} using a gradient descent with a line search. This requires computing the gradient of the total potential with respect to the positions of the particles $i$:
\begin{equation}\label{eq:gradpotential_i}
    \nabla_iV_\text{tot}=\sum_j\nabla_iV_{ij}=2\sum_jh_{ij}\left(\frac{\partial V_{ij}}{\partial d_{ij}}\mathbf{e}_{ij}+\left(2V_{ij}-\frac{d_{ij}}{h_{ij}}\frac{\partial V_{ij}}{\partial d_{ij}}\right)\underbrace{\nabla_ih_{ij}}_{\approx 0}\right)\approx2\sum_jh_{ij}\frac{\partial V_{ij}}{\partial d_{ij}}\mathbf{e}_{ij}\, .
\end{equation}
The unit vector pointing from particle $j$ to particle $i$ is $\mathbf{e}_{ij}=(\mathbf{x}_i-\mathbf{x}_j)/\|\mathbf{x}_i-\mathbf{x}_j\|_2$.
Since our method assumes (and ensures) that the surface is always well sampled, we neglect the contribution of $\nabla_i h_{ij}$, which would otherwise also be difficult to compute because $h_{ij}$ is not differentiable.

We iteratively adjust the particle positions $\mathbf{x}_i$ by moving the particles down their respective gradients over iterations $k$ with step sizes $\gamma^k$,
\begin{equation}\label{eq:incrementrule}
    \mathbf{x}_i^{k+1}=\mathbf{x}_i^k-\gamma^k\nabla_iV_\text{tot}.
\end{equation}
The step length $\gamma^k$ is found by a line search at each iteration $k$. The upper bound of the search interval is given by the Courant--Friedrichs--Lewy (CFL) condition for the stability of the gradient descent \cite{de2013courant}:
\begin{equation}\label{eq:cfl}
    \gamma^k_\text{max} \leq \frac{0.5h_0}{\|\nabla_i V_\text{tot}\|_\infty}\, .
\end{equation}
We here choose the lower bound of the search interval as $\gamma^k_\text{min}=10^{-5}\gamma^k_\text{max}$. If the gradient flow oscillates around local minima, lowering $\gamma^k_\text{min}$ can restore convergence. The gradient descent stops when the following convergence criterion is met over three successive iterations:
\begin{equation}
    \frac{|V_\text{tot}(\mathbf{x}_0^{k+1},\ldots,\mathbf{x}_{n_p-1}^{k+1})-V_\text{tot}(\mathbf{x}_0^k,\ldots,\mathbf{x}_{n_p-1}^k)|}{V_\text{tot}(\mathbf{x}_0^k,\ldots,\mathbf{x}_{n_p-1}^k)}\leq\varepsilon_\text{opt}
\end{equation}
with a user-defined tolerance $\varepsilon_\text{opt}$. This leads to a gradual relaxation of the particle distribution toward a local optimum fulfilling the required curvature adaptivity.

\subsection{Adjusting the size of the particle set}

We assess a lack or excess of particles in the neighborhood of a particle $i$ by computing an integral support measure $S_i$ based on the occupied and expected surface areas of individual particles. A similar concept, but with a geometric estimation of the occupied surface area, previously used local triangulations of particle distributions \cite{suchde2019fully}. 
We here avoid triangulation by direct density summation, which is classic for estimating density fields in Smoothed Particle Hydrodynamics (SPH) \cite{monaghan2005smoothed}. The volume, or in our case surface area, per particle is obtained as the inverse density $\rho_i$. The density $\rho_i$ is computed as a kernel-weighted sum over particles $j$ within the neighborhood radius $r_s$. To account for the locally varying resolution, we modify the density summation slightly and add a weighting factor according to the fraction of the target surface area field $A_{j,\text{theo}}/A_{i,\text{theo}}$, hence
\begin{equation}\label{eq:surfaceareacomputation}
    A_i = 1/\rho_i=1/\left(\sum_jW_{ij}\frac{A_{j,\text{theo}}}{A_{i,\text{theo}}}\right).
\end{equation}
The kernel function $W$ is compact, positive-definite, and normalized. We find that triangular hat functions yield a reasonable estimator and use
\begin{equation}
    W_{ij}=W(d_{ij}, h)=\begin{cases}
        c_{2D}\left(1 - \frac{d_{ij}}{r_s}\right)\quad&\text{if}~\frac{d_{ij}}{r_s}<1\, ,\\
        0,&\text{else}\, ,
    \end{cases}
\end{equation}
with a cutoff radius of $r_s=2h$. 
The scalar constant $c_{2D}=3/(\pi r_s^2)$ is the normalization factor for 2D space, as we are concerned with 2D surfaces embedded in 3D space.
The target surface areas in Eq.~\eqref{eq:surfaceareacomputation} are proportional to the square of the characteristic length,
\begin{equation}\label{eq:targetsurfacearea}
    A_{i,\text{theo}}=ah^2_i \, .
\end{equation}
The constant $a$ depends on the spatial arrangement of the particles. Typically, we encounter either a resulting hexagonal lattice, for which $a=\sqrt{3}/2$, or a square lattice, for which $a=1$. In Eq.~\eqref{eq:surfaceareacomputation}, however, this is inconsequential as $a$ cancels out to yield:
\begin{equation}
    \frac{A_{j,\text{theo}}}{A_{i,\text{theo}}}=\left(\frac{h_j}{h_i}\right)^{\!\!2}.
\end{equation}
The computed occupied surface area is then compared to the theoretical surface area per particle
to obtain the support measure
\begin{equation}
    S_i=\frac{A_{i,\text{theo}}}{A_i}\, .
\end{equation}
Since the value of $a$ cannot be neglected here, we choose the mean of the parameters for hexagonal and square ordering, hence $a\approx0.933$. 

The support measure $S_i$ is used to detect a local lack or excess of particles. Values close to 1 indicate an appropriate number and spacing of particles. Values below 1 indicate a local lack of particles, and values larger than 1 an excess. To render the algorithm robust to oscillations, we introduce hysteresis and act as follows:
\begin{align}
    &\text{if}~S_i<S_\text{min},\quad\text{add particle in the neighborhood,}\label{eq:minsupport}\\
    &\text{if}~S_i>S_\text{max},\quad\text{fuse with a particle in the neighborhood.}\label{eq:maxsupport}
\end{align}
The threshold values $S_\text{min}$ and $S_\text{max}$ determine the feasible set of particle numbers. A smaller interval requires more iterations for the algorithm to converge, whereas a larger interval can cause nearest-neighbor distances to deviate more from the target spacing $h_0$. A finite range is necessary, since the final arrangement of particles, and hence the exact value of $a$, is not known {\it a priori}. Here, we choose the interval so as to allow for at least a $\pm 10\%$ variation in $a$. Specifically, we find $S_\text{min}=0.7$ and $S_\text{max}=1.25$ to work well for the majority of cases.

Fusion removes particle $i$ and its nearest neighbor, and inserts a new particle at the midpoint of their connecting line. Adding a particle is done towards areas of low density $\rho$. For this, we re-use the previously computed densities to approximate the gradient
\begin{equation}\label{eq:densitygradient_i}
    \nabla \rho_i \approx \sum_j\nabla W_{ij}\frac{A_{j,\text{theo}}}{A_{i,\text{theo}}}
\end{equation}
to insert a new particle at position
\begin{equation}\label{eq:particleinsertion}
    x_{\text{new},d}=x_{i,d}-h_i(1+\mu_d)\frac{\left[\nabla\rho_i\right]_d}{\|\nabla\rho_i\|_2}.
\end{equation}
The pseudo-random variables $\mu_d$ are {\it i.i.d.} 
in each spatial dimension $d\in\{0,1,2\}$ from the uniform distribution over the interval $[-0.5,0.5]$, adding independent noise to each component $[\cdot]_d$ of the density gradient. 
We empirically find that this randomness accelerates convergence of the algorithm. This is to be expected, since the optimization problem is highly non-convex, so that stochasticity helps escape local minima. We also find that the randomness helps avoid oscillations in which particles are added and then immediately removed again.

The algorithm inserts and fuses particles in a loop with iteration counter $k_\pm$ until the relative change in the particle number is smaller than a threshold $\varepsilon_{n_p}$,
\begin{equation}\label{eq:convergencesize}
    \frac{|n_{p,\mathrm{new}}-n_{p,\mathrm{old}}|}{n_{p,\mathrm{old}}}\leq\varepsilon_{n_p}.
\end{equation}
In the results presented below, we use the same stopping tolerance for the gradient descent and for the particle number, $\varepsilon_\text{opt}=\varepsilon_{n_p}=:\varepsilon_\text{SAISS}$. This tolerance is the main user-adjustable parameters of the method and determines the trade-off between the quality of the discretization and the required computational effort. More complex surfaces with larger variation in curvature, as well as problems with higher curvature adaptivity, require smaller tolerances. When tuning the tolerance, the quality of the resulting point distribution can be monitored by tracking deviations from the prescribed nearest-neighbor distances.

\subsection{Optimization within a curved surface}

Since the space in which we perform the optimization is curved, we pay special attention to the computation of positions, vectors, and distances. 
As geodesic distance computations on arbitrary curved surfaces are computationally expensive, we approximate local geodesic distances $d_g$ within particle neighborhoods by Euclidean distances $d_e$ in the embedding space. This is a valid approximation, since the algorithm drives the discretization to well-sample the surface. If the sampling condition
\begin{equation}
    d_g\leq\frac{\pi}{\kappa_\text{max}}
\end{equation}
is fulfilled for a local geodesic with bounded absolute principal curvature $\kappa_\text{max}$, the approximation of $d_g$ by the Euclidean distance $d_e$ is bounded from above and below by \cite{bernstein2000graph}
\begin{equation}
    d_g-\frac{d_g^3}{24}\kappa_\text{max}^2\leq d_e\leq d_g.
\end{equation}
While Euclidean distance consistently underestimates geodesic distance, the approximation error converges cubically with the geodesic distance between the two points. Since in our case $d_g \in O(\tilde{h})$, the error introduced by using Euclidean distances converges with order three in the resolution $\tilde{h}$.

Vector fields on curved surfaces have to be correctly projected into the tangent space of the surface. We therefore project the displacements in Eq.~\eqref{eq:gradpotential_i} and the density gradients in Eq.~\eqref{eq:densitygradient_i} as:
\begin{equation}\label{eq:tangentprojection}
    \mathbf{v}_i=\left(\mathbf{I}-\mathbf{n}_i\otimes\mathbf{n}_i\right)\hat{\mathbf{v}}_i\, ,
\end{equation}
in which $\hat{\mathbf{v}}_i$ and $\mathbf{v}_i$ denote any vector-valued quantity on particle $i$ before and after projection, respectively, and $\mathbf{n}_i$ the unit surface normal.
Additionally, we ensure that all particles lie within the surface at all times by projecting them onto their respective closest point on the surface after every change to their position,
\begin{equation}\label{eq:cptransform}
    \mathbf{x}_{i}=\mathbf{cp}(\hat{\mathbf{x}}_i)\, .
\end{equation}
Specifically, this closest-point transform is applied after moving particles down the potential gradient (Eq.~\eqref{eq:incrementrule}), during and after the line search of the gradient descent, after fusing particles, and after adding a particle. 

Computing these projections requires geometric information about the surface at the location of each particle: the local surface normal, the closest point on the surface, and---for curvature adaptivity according to Eq.~\eqref{eq:htilde}---the principal curvatures. Since we assume the surface to be implicitly defined by a level-set, we accurately and efficiently determine these geometric quantities using the PCP algorithm \cite{schulze2024high}. The PCP algorithm locally approximates the level-set function by polynomials, from which geometric quantities can be computed exactly. 
The polynomial representation of the surface guarantees globally consistent curvature values and readily determines the closest point on the surface. The closest-point transform is often avoided in the literature by introducing normal forces attracting the particles toward the surface \cite{de1992physically}. Here, we directly impose that particles remain on the zero level-set up to solver tolerance. 

Throughout this paper, we use polynomials of $L_1$-degree four. This guarantees that closest points are computed to fifth order, surface normals with order four, and curvatures with an order of convergence of three. An inherent limitation of our approach, however, is the fact that smooth polynomials cannot represent sharp corners or edges in the surface. At those critical points, the accuracy of our approach falls back to a first-order approximation with respect to the resolution of the level-set discretization.

\subsection{Overall algorithm}

\begin{figure}
    \centering
    \tikzstyle{block} = [rectangle, draw, text width=15em, text centered, rounded corners, minimum height=3em]
\scalebox{0.8}{\begin{tikzpicture}
 [node distance=1.5cm]
\node (n1) at (0,0) [block]  {Initialize particle distribution $\mathbf{x}_i^0$, compute global potential};
\node (n2) [block, below of=n1, yshift=-0.5cm] {\textbf{while} \it{relaxation}};
\node (n3) [block, below of=n2, xshift=4cm] {\textbf{while} \it{finding size}};
\node (n4) [block, below of=n3, xshift=4cm] {Add particles};
\node (n5) [block, below of=n4, yshift=-0.5cm] {Fuse particles};
\node (n6) [block, below of=n5, yshift=-0.5cm] {Check convergence \it{finding size}};
\node (n7) [block, below of=n6, xshift=-4cm] {Compute potential gradient};
\node (n11) [block, below of=n7, yshift=-0.5cm] {Perform line search: \textbf{while} $\gamma<\gamma_\text{max}~\&~V_\text{tot} < V_\text{tot,old}$};
\node (n12) [block, below of=n11, xshift=4cm] {Update particle positions};
\node (n8) [block, below of=n12, yshift=-0.5cm] {Update global potential, $\gamma$};
\node (n9) [block, below of=n8, xshift=-4cm] {Check convergence \it{relaxation}};
\node (n10) [block, below of=n9, xshift=-4cm] {Return $\mathbf{x}^k_i$};
\draw [->] (n1) -- (n2);
\draw [->] (n4) -- node [right]{Call gcf, project added particles onto surface}(n5);
\draw [->] (n5) -- node [right]{Call gcf, project fused particles onto surface}(n6);
\draw [->] (n7) -- node [right]{Project into tangent space, initialize $\gamma=\gamma_\text{min}$} (n11);
\draw [->] (n12) -- node [right]{Call gcf, project moved particles onto surface} (n8);
\draw [->] (n11.south) -| ++(0,-1) |- (n12.west);
\draw [->] (n2.south) -| ++(0,-1) |- (n3.west);
\draw [->] (n3.south) -| ++(0,-1) |- (n4.west);
\draw [->] (n6.south) -| ++(0,-1) |- (n7.east);
\draw [-] (n6.west) -| ++(-4.25,0) |- (n3.west);
\draw [->] (n8.south) -| ++(0,-1) |- (n9.east);
\draw [->] (n8.west) -| ++(-4.25,0) |- (n11.west);
\draw [->] (n9.south) -| ++(0,-1) |- (n10.east);
\draw [->] (n9.west) -| ++(-4.25,0) |- (n2.west);
\end{tikzpicture}
}    
    \caption{The SAISS algorithm for finding a locally regular and globally curvature-adaptive particle discretization of an implicit curved surface. The abbreviation ``gcf'' stands for ``geometric computing framework'', which in our case is the particle closest-point (PCP) method, that computes surface normals, principal curvatures, and closest points on the surface.}\label{fig:algorithm}
\end{figure}
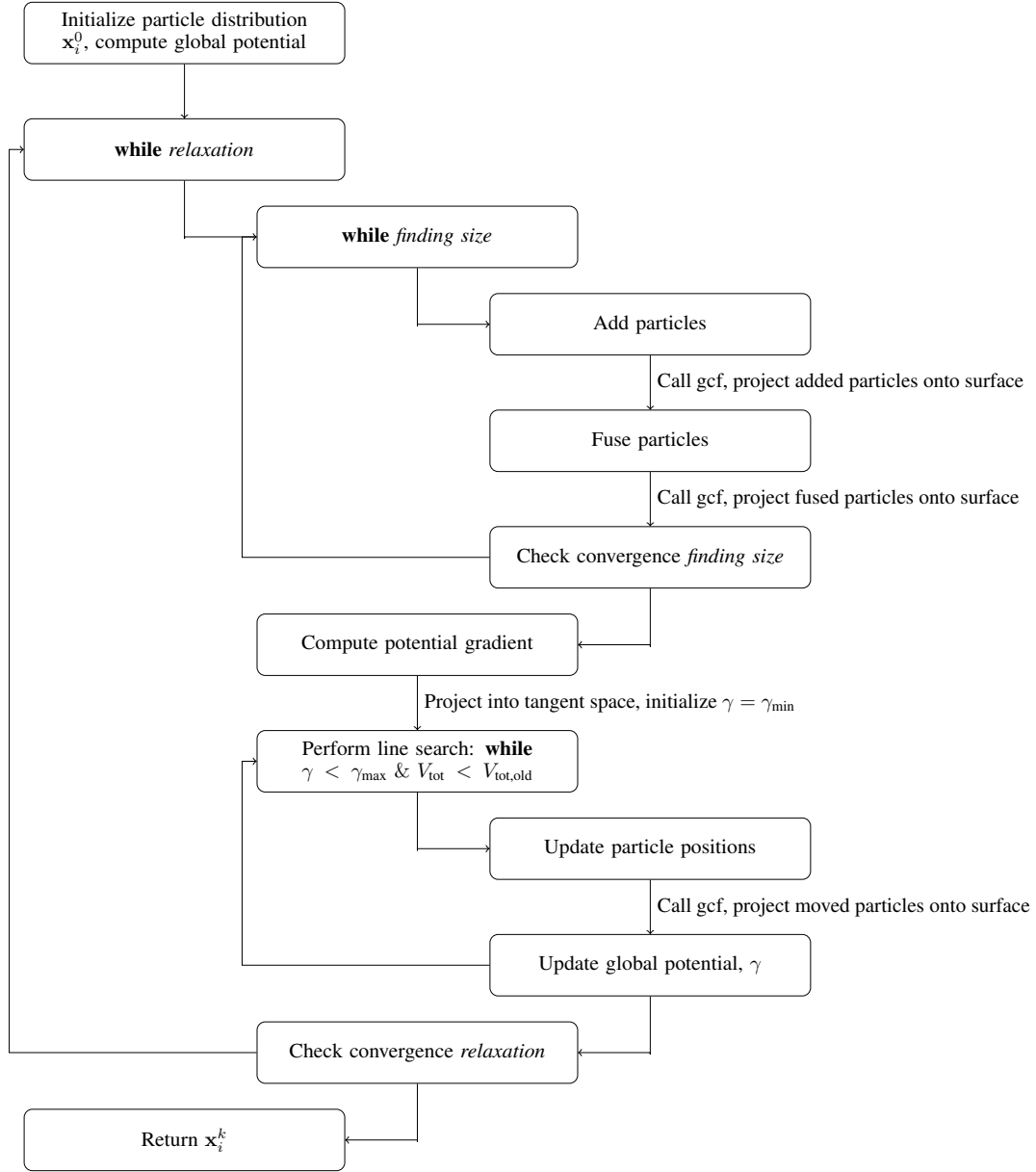

In the final SAISS algorithm, the two steps of finding the particle set size and relaxing the particle positions are nested as shown in the flow chart in Fig.~\ref{fig:algorithm}. This nesting avoids the need for the particle distribution to be perfectly regular after relaxation. Since addition and fusion of particles limits the step size for the relaxation, we fix the particle number $n_p$ and skip the size-finding routine as soon as no particles were fused or added during three subsequent iterations of the outer loop.

\section{Benchmarks and Results}
We test the SAISS algorithm to compute adaptive particle discretizations with different resolutions for a variety of parametric and non-parametric curved surfaces. Since analytical solutions for optimal particle distributions are rarely available, we assess errors by comparing nearest-neighbor distances to the local target discretization length. We further showcase the application of the SAISS method to computing surface integrals by determining the total surface area of the shapes.

\subsection{Parametric curved surfaces}

We validate the point distributions computed by the SAISS algorithm on parametric surfaces, where geometric quantities are analytically known. We start with the simplest case, the sphere, for which the globally optimal arrangement of points is known, providing a direct validation. We then benchmark the curvature adaptivity on ellipsoids of different aspect ratio before showing applications to a locally deformed surface sheet and a triply periodic minimal surface (TPMS).

\subsubsection{3D sphere}

The 3D unit sphere is defined as the zero level-set of the function
\begin{equation}
    \phi(\mathbf{x})=\sqrt{\frac{x^2}{A^2}+\frac{y^2}{B^2}+\frac{z^2}{C^2}}-1,
\end{equation}
with $A=B=C=1$. To leverage the geometric description of the PCP method, we initialize bulk particles in a narrow band of width $w=10h_b$ around the surface at locations $\mathbf{x}_p$ with level-set function values $\phi(\mathbf{x}_p)$. The particles are initialized in a Cartesian grid with inter-particle spacing $h_b=1/16$. Using the PCP method, we compute a representation of the surface using local polynomials obtained in neighborhoods with a cutoff radius of $r_c=2.4h_b$. We store the geometric representation in an {\tt OpenFPM} distributed vector data structure, containing the sample particles with their respective polynomial coefficients as particle properties. 

We then start the SAISS method from an initial surface particle distribution given by the sample particles of the PCP method. For SAISS, we use a cutoff radius of $r^*=2.5h$ for the inter-particle potential and a cutoff radius of $r_s=2h$ for approximating the local density. We use the same solver tolerance for the PCP method and the SAISS method, setting both to $\varepsilon_\text{PCP}=\varepsilon_\text{SAISS}=10^{-10}$. We choose a very small tolerance here to demonstrate the achievable accuracy of the SAISS method. In practice, larger tolerances are often sufficient. For adding and fusing particles, we use lower and upper support bounds of $S_\text{min}=1$ and $S_\text{max}=1.25$. Since the sphere is a surface of constant curvature, the curvature-adaptivity coefficient and the reference curvature are $\tau=\kappa_\text{ref}=0$, and we only adapt the characteristic length $h_0$.

\begin{figure}
    \centering
    \begin{subfigure}{\linewidth}
        \centering
        \includegraphics[width=0.5\linewidth]{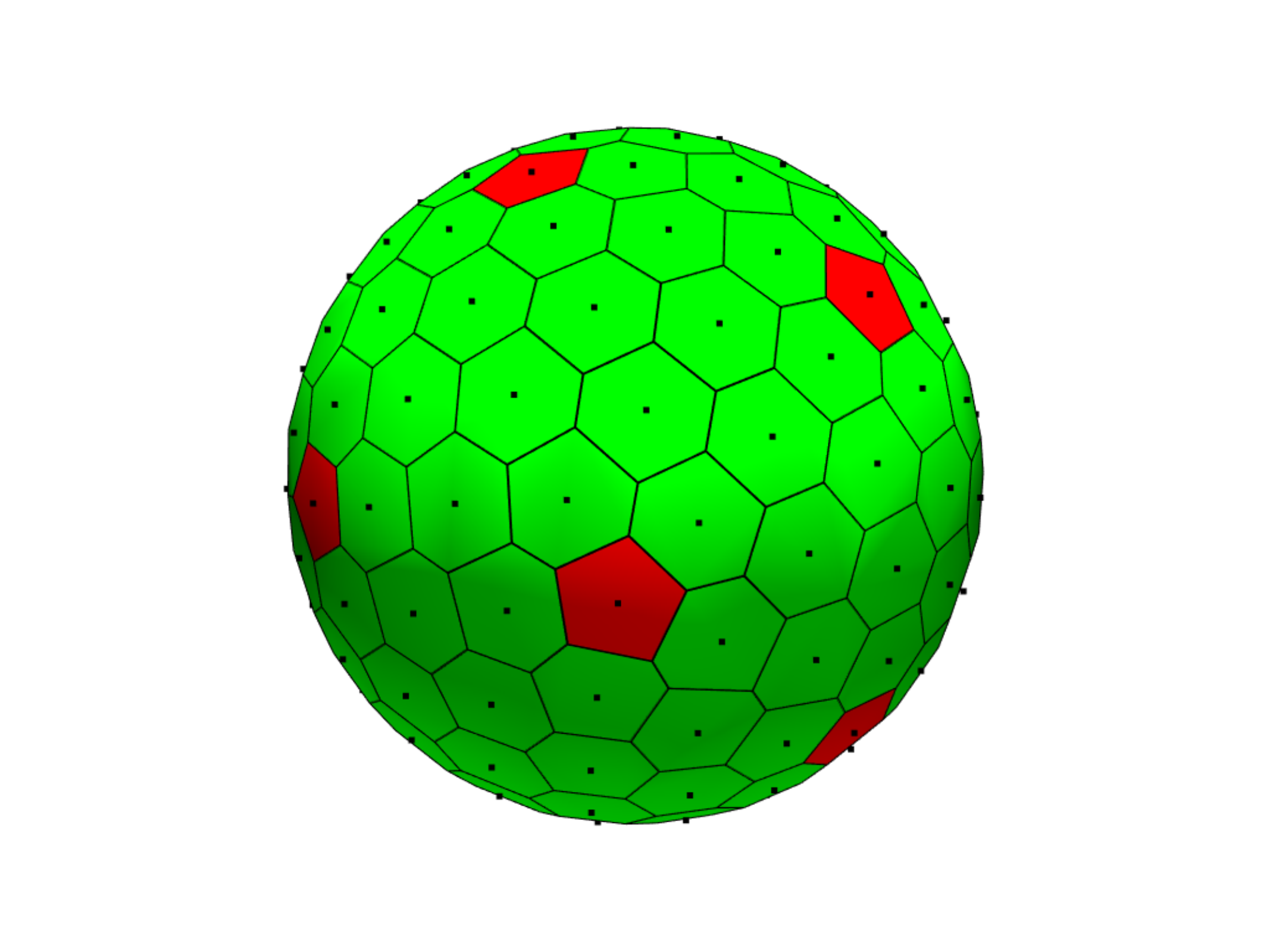} 
        \caption{}\label{fig:spherevoronoicoarse}
    \end{subfigure}
    \vspace{1em}
    \begin{subfigure}{\linewidth}
        \centering
        \includegraphics[width=0.95\linewidth]{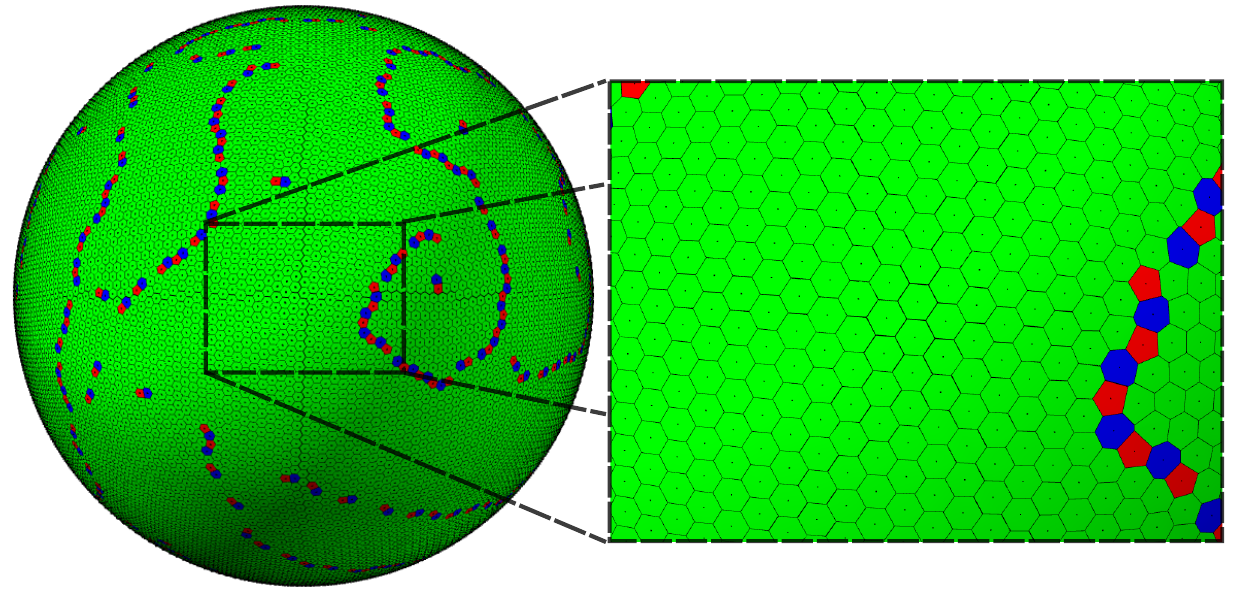}
        \caption{}\label{fig:spherevoronoifine}
    \end{subfigure}
    \caption{Voronoi tessellation of the particle distribution computed using the SAISS method for $h_0=0.3$ (a) and $h_0=0.03$ (b). The dots are the particle positions, which are used as generators for the associated Voronoi cells. Green cells represent hexagons, red cells are pentagons, and blue cells are heptagons.}
    \label{fig:spherevoronoi}
\end{figure}

The results are shown in Fig.~\ref{fig:spherevoronoi} for $h_0=0.3$ (a) and $h_0=0.03$ (b).
We assess the quality of the resulting particle distribution by computing a Voronoi tessellation \cite{virtanen2020scipy,caroli2009robust} with particles as the generator points for Voronoi cells. Since there is a hysteresis in the support thresholds $S_\text{min}$ and $S_\text{max}$, and the characteristic length is chosen to be the minimum of any two particles ($h_{ij}=\min(h_i,h_j)$), the number of particles is bound to be larger than the theoretically required minimum\footnote{A slightly larger number of particles is desirable in downstream applications in which a certain minimum support in a neighborhood is required to yield full-rank matrices, e.g., for solving linear systems of equations.}. Therefore, the particle arrangement tends toward the most efficient packing, which is the hexagonal packing. It is, however, not possible to tile a sphere using hexagons only. If we tile the sphere in a globally optimal configuration with hexagons and pentagons, exactly 12 pentagons are required. 

We count the number of polyhedrons that are generated in the Voronoi tessellation and find 154 hexagons and 12 pentagons for $h_0=0.3$, indicating that the SAISS method has converged to the globally optimal arrangement of particles. We consistently observe this behavior for larger $h_0$ and hence smaller numbers of particles $n_p$. For small $h_0$, the Voronoi tessellation includes pentagons and heptagons in addition to hexagons. For $h_0=0.03$, we find 15\,570 hexagons, 425 pentagons, and 413 heptagons. As seen in Fig.~\ref{fig:spherevoronoifine}, the pentagons and heptagons are mainly arranged in lines wrapping around the sphere. We hypothesize that the exponential growth in the number of local minima with increasing $n_p$ leads the algorithm to converge to a local optimum. The globally optimal configuration would have no heptagons, 12 pentagons, and otherwise hexagons. Nevertheless, the SAISS method yields about 95\% hexagons and generally a good agreement between the target and actual nearest-neighbor distances with an average relative error $|h_{ij}-h_0|/h_0=0.0381$. This average error is a conservative assessment of accuracy, as it does not account for the bias toward larger particle numbers. This bias inevitably leads to average nearest-neighbor distances smaller than the target.

We use the fact that an analytical expression is available for the surface area of the 3D unit sphere $\Gamma$, $A=4\pi$, to judge the quality of the final surface discretization when computing the surface integral 
\begin{equation}
    A=\int_\Gamma\mathrm{d}\Gamma\approx\sum_iw_i=\sum_iA_i \, ,
\end{equation}
where the quadrature weights $w_i$ are the individual surface areas $A_i$ computed according to Eq.~\eqref{eq:surfaceareacomputation}. We use SAISS to find surface particle distributions for $h_0\in\{0.015,0.03,0.06,0.12,0.24\}$ compute the respective surface integrals $A_h$. We find relative errors $e(A)=(4\pi-A_h)/4\pi$ of $\{0.0052,0.0055,0.0053,0.008,0.013\}$. 

\subsubsection{3D ellipsoid}

The non-constant curvature of the ellipsoid allows us to test the capability of the SAISS method to compute curvature-adaptive particle distributions. We consider two different rotationally symmetric ellipsoids: (1) $A=0.75,\,B=C=0.5$, and a version with higher maximum curvature, (2) $A=0.75,\, B=C=0.125$. For both ellipsoids, we compute an initial surface particle distribution using the same procedure as for the 3D sphere, with a bulk inter-particle spacing of $h_b=1/32$ for ellipsoid (1) and $h_b=1/128$ for ellipsoid (2). This results in an initial particle set size of $n_p=5264$ for ellipsoid (1) and $n_p=18\,680$ for ellipsoid (2). 

\begin{figure}
    \centering
    \includegraphics[width=\linewidth]{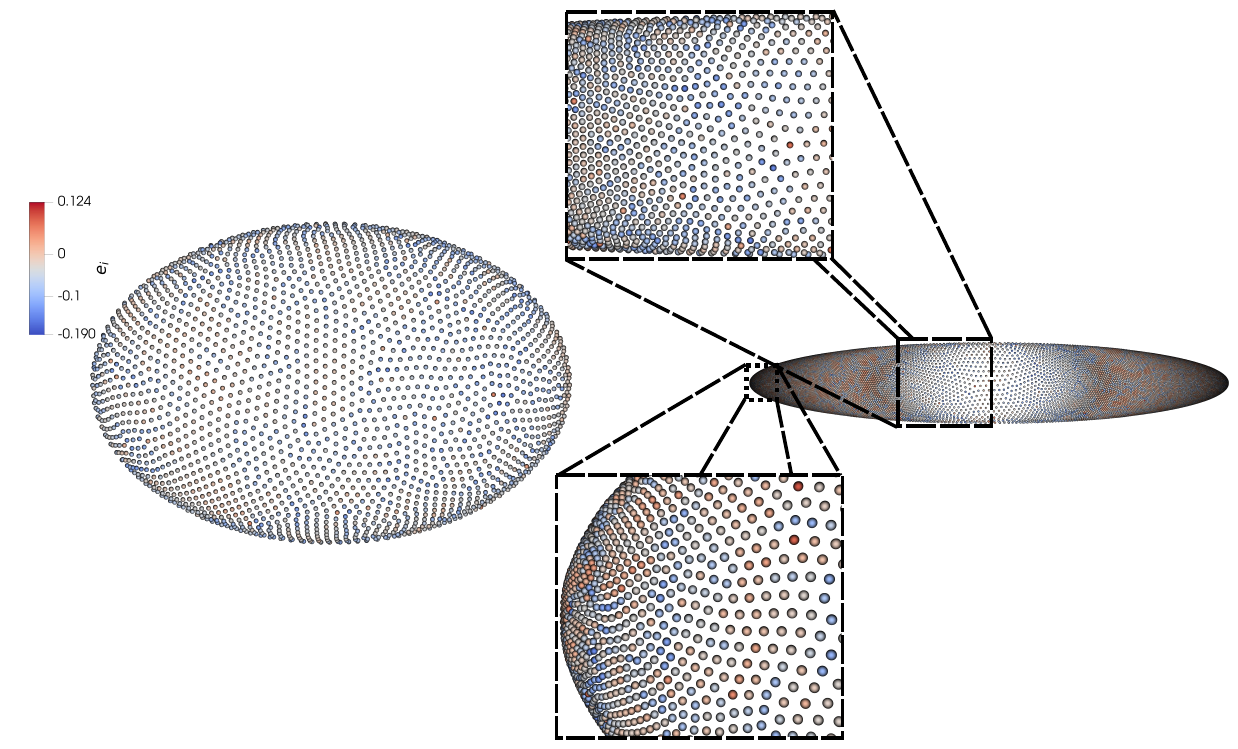}
    \caption{Computed particle discretization on ellipsoid (1) with $\tau=0$ and $\varepsilon_\text{SAISS}=10^{-4}$ (left) and on ellipsoid (2) with $\tau=5$ and $\varepsilon_\text{SAISS}=10^{-5}$ (right). The particles are color-coded according to their respective relative errors $e_i=(d_{\min}-h_i)/h_i$ (color bar). The insets for ellipsoid (2) show areas of particularly high and low curvature to illustrate the curvature adaptivity of the particle distribution.}
    \label{fig:3DellipsoidParticleDistributions}
\end{figure}

As parameters for the SAISS method, we use $r^*=2h$, $r_s=2h$, $S_\text{min}=0.8$, and $S_\text{max}=1.25$. We set the target spacing to $h_0=1/30$ for ellipsoid (1) and to $h_0=1/60$ for ellipsoid (2), assumed at the points of lowest deviation from flatness, i.e., where $x=0$. At these points, the principal curvatures for ellipsoid (1) are $\kappa_1=2$ and $\kappa_2=8/9$, such that we set the reference curvature to $\kappa_\text{ref}=\sqrt{2^2+\frac{8}{9}^2}$. For ellipsoid (2), the principal curvatures are $\kappa_1=8$ and $\kappa_2=2/9$, motivating $\kappa_\text{ref}=\sqrt{8^2+\frac{2}{9}^2}$.
We vary the curvature-adaptivity parameter $\tau$ and the tolerance $\varepsilon_\text{SAISS}$ and compute results for each parameter sets as reported in Table~\ref{tab:SAISSellipsoid}. Fig.~\ref{fig:3DellipsoidParticleDistributions} shows exemplary final particle distributions on the two ellipsoids.

Depending on the curvature-adaptivity parameter $\tau$, we find particle discretizations with smaller, similar, and larger numbers of particles than the initial condition. We find an up to 5-fold downsampling for ellipsoid (2) with $\tau=0$ and a 15-fold upsampling for ellipsoid (1) with $\tau=50$. In all cases, the final particle set sizes $n_{p,\text{final}}$ are determined robustly in few 
iterations $k_{n_p}$ of the optimization loop, see Table~\ref{tab:SAISSellipsoid}. The most iterations were required for ellipsoid (1) with the stricter tolerance and the maximum curvature-adaptivity, $\tau=50$, where finding the number of particles required $k_{n_p}=26$ iterations of the gradient descent. For the same ellipsoid and $\tau$, we observe a consistent improvement both in the $L_2$- and $L_\infty$ errors of the nearest-neighbor distances when the tolerance is reduced from $\varepsilon_\text{SAISS}=10^{-4}$ to $\varepsilon_\text{SAISS}=10^{-5}$. This comes at the cost of requiring more gradient-descent iterations $k$ until convergence of the overall optimization problem at $k=k_\text{opt}$. The higher-curvature ellipsoid (2) also requires more iterations until convergence than ellipsoid (1). The final errors for ellipsoid (1) and (2) are comparable.

\begin{table}[]
    \centering
    \begin{tabular}{ l|c|c|c|c|c }
    Parameter set & $k_{n_p}$ & $k_\text{opt}$ & $n_{p,\text{final}}$ & $\|e\|_\infty$ & $\|e\|_2$\\ \hline
    ellipsoid (1), $\varepsilon_\text{SAISS}=10^{-4}$, $\tau=0$ & 1 & 109 & 4304 & 0.14 & 0.048 \\  
    ellipsoid (1), $\varepsilon_\text{SAISS}=10^{-4}$, $\tau=0.5$ & 1 & 141 & 5062 & 0.13 & 0.039 \\
    ellipsoid (1), $\varepsilon_\text{SAISS}=10^{-4}$, $\tau=5$ & 8 & 76 & 12\,061 & 0.14 & 0.035 \\
    ellipsoid (1), $\varepsilon_\text{SAISS}=10^{-4}$, $\tau=25$ & 12 & 86 & 43\,566 & 0.18 & 0.031 \\
    ellipsoid (1), $\varepsilon_\text{SAISS}=10^{-5}$, $\tau=25$ & 25 & 155 & 43\,450 & 0.15 & 0.028 \\
    ellipsoid (1), $\varepsilon_\text{SAISS}=10^{-4}$, $\tau=50$ & 15 & 63 & 82\,196 & 0.27 & 0.028 \\
    ellipsoid (1), $\varepsilon_\text{SAISS}=10^{-5}$, $\tau=50$ & 26 & 139 & 82\,398 & 0.21 & 0.027 \\ \hline
    ellipsoid (2), $\varepsilon_\text{SAISS}=10^{-4}$, $\tau=0$ & 5 & 88 & 3788 & 0.15 & 0.048 \\
    ellipsoid (2), $\varepsilon_\text{SAISS}=10^{-4}$, $\tau=0.5$ & 10 & 95 & 7642 & 0.20 & 0.043 \\
    ellipsoid (2), $\varepsilon_\text{SAISS}=10^{-5}$, $\tau=0.5$ & 6 & 267 & 7636 & 0.19 & 0.034 \\
    ellipsoid (2), $\varepsilon_\text{SAISS}=10^{-4}$, $\tau=1$ & 7 & 92 & 11\,472 & 0.22 & 0.041 \\
    ellipsoid (2), $\varepsilon_\text{SAISS}=10^{-5}$, $\tau=1$ & 9 & 259 & 11\,471 & 0.18 & 0.033 \\
    ellipsoid (2), $\varepsilon_\text{SAISS}=10^{-5}$, $\tau=5$ & 19 & 145 & 41\,982 & 0.19 & 0.030 \\
    ellipsoid (2), $\varepsilon_\text{SAISS}=10^{-5}$, $\tau=10$ & 16 & 260 & 79\,673 & 0.18 & 0.026
    \end{tabular}\vspace{1em}\caption{Number of gradient-descent iterations until the size of the particle set is determined ($k_{n_p}$), number of gradient-descent iterations to solve the overall optimization problem ($k_\text{opt}$), final size of the particle set ($n_{p,\text{final}}$), and maximum and mean errors ($\|e\|_\infty$, $\|e\|_2$) in the nearest-neighbor spacing for different ellipsoids, solver tolerances ($\varepsilon_\text{SAISS}$), and curvature-adaptivity parameters ($\tau$).}\label{tab:SAISSellipsoid}
\end{table}

\begin{figure}
    \centering
    \def\svgwidth{0.7\textwidth}
    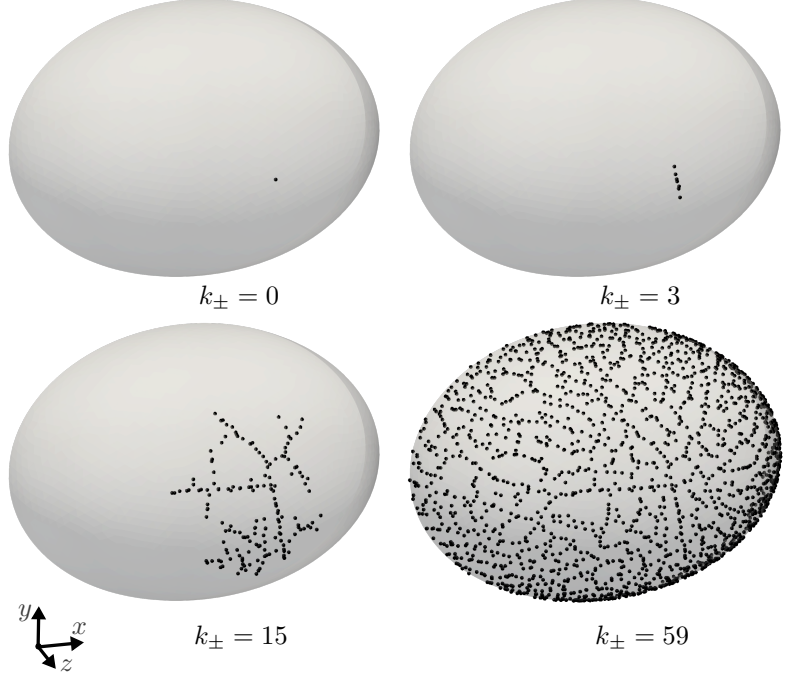
    \caption{Visualization of the particle distribution on ellipsoid (1) when starting from a single particle at iteration $k=0$. The ellipsoid is clipped along the $x-y$ plane for better visualization. After $k_\pm=59$ iterations of particle fusion and addition, the entire ellipsoid is covered by particles of sufficient density, which subsequently relax to the regular distribution shown in Fig.~\ref{fig:3DellipsoidParticleDistributions}.}
    \label{fig:3DellipsoidExploration}
\end{figure}

We further use this test case to visualize the behavior of the particle insertion routine. For this, instead of using the PCP sample particle distribution as initial condition, we start from a single particle. From there, as shown in Fig.~\ref{fig:3DellipsoidExploration}, the SAISS algorithm explores the entire surface. This is done by expanding the particle set along lines ($k_\pm=3$) that subsequently branch ($k_\pm=15$) until the conditions in Eqs.~\eqref{eq:minsupport} and \eqref{eq:maxsupport} are fulfilled for all particles ($k_\pm=59$). At this point, the surface is well-covered, and the distribution is fit for subsequent relaxation to the regular state shown in Fig.~\ref{fig:3DellipsoidParticleDistributions}.

To motivate our choice of using an integral support measure to diagnose a local lack or excess of particles, we compare our approach with the traditional approach of using the number of neighbors and their minimum distance as criteria \cite{reboux2012self}. There, particles are inserted whenever a minimum number of neighbors within the cutoff radius is not found, and a pair of particles is fused whenever their distance is less than $h_i/2$.
Using this alternative approach on ellipsoid (1) with the same neighborhood radius $r_s=2h$ and a minimum of 10 required neighbors ($\varepsilon_\text{SAISS}=10^{-4}$, $\tau=0$) results in $n_{p,\text{final}}=4254$, which is comparable to the number of particles obtained by SAISS. The two approaches also yield final particle distributions of similar quality with a maximum absolute value of the relative error in spacing of $\|e\|_\infty=0.14$ (same as SAISS) and an average of $\|e\|_2=0.044$ (SAISS: $\|e\|_2=0.048$).  Inserting and fusing particles depending on the number of particles in a neighborhood and the nearest neighbor distance, however, slows down convergence of the algorithm ($k_{n_p}=9$ instead of $k_{n_p}=1$ and $k_\text{opt}=126$ instead of $k_\text{opt}=109$). 

The advantage of the present integral support measure becomes even more evident for curvature-adaptive particle distributions. For the last parameter set on ellipsoid (2) ($\varepsilon_\text{SAISS}=10^{-5}$, $\tau=10$), the classic algorithm does not converge. It keeps adding and fusing particles until the gradient descent stops after $k_\text{opt}=319$ steps oscillating around a local minimum. This eventually yields a particle distribution with $n_{p,\text{final}}=91\,064$ particles spaced too densely, with an average absolute value of the relative error of $\|e\|_2=0.062$ (SAISS: $\|e\|_2=0.026$) and a maximum of $\|e\|_\infty=0.48$ (SAISS: $\|e\|_\infty=0.18$). The integral support measure effectively avoids oversampling areas of large resolution gradients by weighting the particle contributions (Eq.~\eqref{eq:surfaceareacomputation}). This represents a continuous field, in contrast to discrete particle counts, which enables convergence of the gradient descent.

As a final test, we assess the run-to-run variability in the determined particle set size $n_{p,\text{final}}$ by performing 20 independent repetitions of each of the two most extreme cases (ellipsoid (1), $\varepsilon_\text{SAISS}=10^{-4}$, $\tau=0$ and ellipsoid (2), $\varepsilon_\text{SAISS}=10^{-5}$, $\tau=10$). The summary statistics are reported in Table~\ref{tab:finalsetsize}. The particle set-size finding routine is robust against its stochastic component: Repeating the computations with different seeds for the random insertion offsets changes the resulting particle set sizes only by a small margin, with a coefficient of variation of CV$=0.15\%$ for the first parameter set and CV$=0.11\%$ for the last parameter set.

\begin{table}[]
    \centering
    \begin{tabular}{ l|c|c|c|c|c }
    Parameter set & mean$(n_{p,\text{final}})$ & $\sigma$ [particles] & CV [\%] & min$(n_{p,\text{final}})$ & max$(n_{p,\text{final}})$ \\ \hline
    ellipsoid (1), $\varepsilon_\text{SAISS}=10^{-4}$, $\tau=0$ & 4310 & 6.39 & 0.15 & 4297 & 4318 \\
    ellipsoid (2), $\varepsilon_\text{SAISS}=10^{-5}$, $\tau=10$ & 79472 & 83.66 & 0.11 & 79334 & 79673 \\
    \end{tabular}\vspace{1em}
    \caption{Summary statistics of the computed final set sizes for the two most extreme test cases, each repeated 20 times. The standard deviation in the number of particles is denoted by $\sigma$, whereas the coefficient of variation is denoted by $\text{CV}=100\sigma/\text{mean}(n_{p,\text{final}})$.}\label{tab:finalsetsize}
\end{table}

\subsubsection{3D Gaussian bump}

We next consider a test case of a surface described explicitly by a truncated Gaussian superimposed onto a plane \cite{bachini2024diffusion}. The surface in Cartesian coordinates is given by the following elevation field over the $x-y$ plane: 
\begin{equation}
\begin{cases}
z=G(x,y)=\exp\left(\frac{-1}{1-d^2}\right)\quad&\mathrm{if}~d<1-\delta\\
0&\mathrm{else,}
\end{cases}
\end{equation}
with $d=\|\mathbf{x}-\mathbf{x}_c\|_2/R$ the normalized distance between the argument $\mathbf{x}=(x,y)$ and the center of the Gaussian bump $\mathbf{x}_c=(x_c,y_c)=(-0.5,0)$. The radius of the bump is $R=0.25$, and $\delta=0.025$ is the threshold for the truncation where the Gaussian transitions to the plane. We consider the domain $[-1,1]^3$ and impose periodic boundary conditions in the $x$- and $y$-directions. While it is straightforward to sample this surface uniformly over $x-y$, doing so in-plane is not trivial. Curvature adaptivity is also challenging, as the curvature increases rapidly in a circular ring around the bottom of the bump.

We transform this surface to an implicit representation by considering the level-set function $\phi(x,y,z)=G(x,y)-z$. We create a narrow-band discretization of the level-set function by placing points on a Cartesian grid with inter-particle spacing $h_b=0.005$ in a tubular neighborhood of width $w=7.5h_b$ around the surface. All parameters of the PCP method remain the same as in the sphere and ellipsoid cases. The PCP method quantifies the surface with mean curvatures roughly between -75 and +25, and Gaussian curvatures roughly between -160 and +140, which is consistent with Ref.~\cite{bachini2024diffusion}. 

For the SAISS method, we set $h_0=0.025$, $S_\text{min}=0.7$, $S_\text{max}=1.25$, $\kappa_\text{ref}=0$, and $r^*=r_s=2h$, and we use a tolerance of $\varepsilon_\text{SAISS}=10^{-7}$. 
As in the previous examples, we start from the PCP sample particles. However, since the target discretization length is a factor of 5 larger than the characteristic length in the sample particle set, we create the initial point set for SAISS by randomly choosing one third of the sample particles (uniform Bernoulli thinning). 

\begin{figure}
    \centering
    \def\svgwidth{0.55\textwidth}
    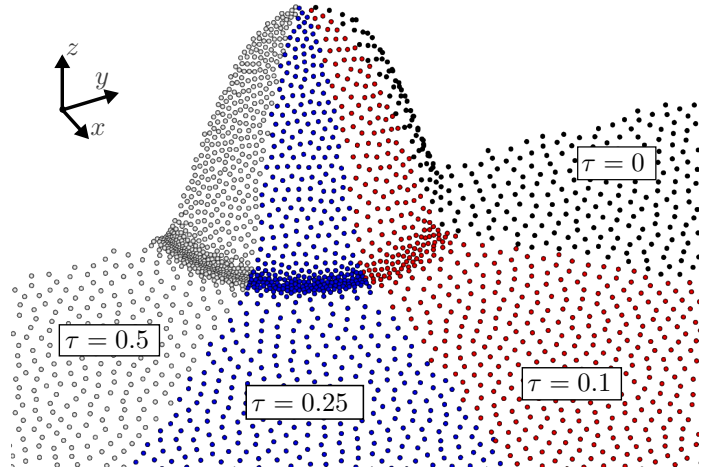
    \caption{Excerpt of the final particle discretizations found by SAISS for the 3D Gaussian bump surface. The visualization focuses on the vicinity of the bump, clipped at the $y-z$ plane for better visualization. The particle distributions computed for $\tau=0$, $\tau=0.1$, $\tau=0.25$, and $\tau=0.5$ are shown in different wedges of this radially symmetric problem and in different colors (black, red, blue and white, respectively).}
    \label{fig:3Dbumpdiscretization}
\end{figure}

Fig.~\ref{fig:3Dbumpdiscretization} shows the resulting particle distributions from SAISS for different values of the curvature-adaptivity coefficient $\tau$. The resulting particle numbers are $n_p=6551$ for $\tau=0$, $n_p=7457$ for $\tau=0.1$, $n_p=8517$ for $\tau=0.25$, and $n_p=10\,227$ for $\tau=0.5$. For all $\tau$, the algorithm required between 3 ($\tau=0$) and 43 ($\tau=0.5$) iterations in the gradient descent to converge to the final particle set size, and a maximum of 50 iterations for the overall optimization problem including the particle equilibration.
The $L2$-norm of the deviation from $h_i$ ranges between 0.045 and 0.066, indicating sufficient local regularity for downstream applications. In the high-curvature region around the rim of the Gaussian bump, however, the bias of the method toward smaller particle spacings is clearly visible: The maximum absolute deviation from the characteristic length $h_i$ is always at a rim particle with neighbor distances up to two-fold smaller than $h_i$.

An inherent trade-off of the SAISS method, that is nicely exposed by the diverging curvature regions of this test case, is the compromise between preventing the gradient descent from converging prematurely to a local minimum and enabling it to converge rapidly. This trade-off is mainly controlled by the solver tolerance, for which we find $\varepsilon_\text{SAISS}=10^{-7}$ to provide good results for the present case. 
The trade-off is also influenced by the bounds of the step-size interval for the line search in gradient descent. Here, we find that a lower minimum step size of $\gamma_\text{min} = 10^{-7}\gamma_\text{max}$ enables faster convergence to particle sets with average deviation from $h_i$ around 5\%, as reported above.

\subsection{TPMS gyroid}

The triply periodic minimal surface (TPMS) gyroid is a smooth curved surface important to many engineering applications. Due to it being a symmetric and minimal surface with a constant mean curvature of zero, it distributes mechanical stresses almost isotropically and has maximum surface-to-volume ratio, which is beneficial in applications involving heat transfer \cite{Petrasch2008b,Petrasch2008}.

Specifically, we consider the truncated TPMS gyroid described by the zero level-set of the function
\begin{equation}
    \phi(\mathbf{x})=\mathrm{cos}(qx)\mathrm{sin}(qy)+\mathrm{cos}(qy)\mathrm{sin}(qz)+\mathrm{cos}(qz)\mathrm{sin}(qx)
\end{equation}
with wavenumber $q=2\pi/\lambda$, and wavelength $\lambda=1$. We discretize the implicitly defined surface in the domain $[-1,1]^3$ and impose periodic boundary conditions in the $x$-, $y$- and $z$-directions.

We compute a representation of the surface using the PCP method in a narrow-band of $w=10h_b$ with $h_b=1/64$ and all other parameters as in the previous cases. For SAISS, we impose a surface resolution $h_0=1/50$ and a solver tolerance of $\varepsilon_\text{SAISS}=10^{-5}$. All other SAISS parameters are the same as in the Gaussian bump case. 

\begin{figure}
    \centering
    \def\svgwidth{\textwidth}
    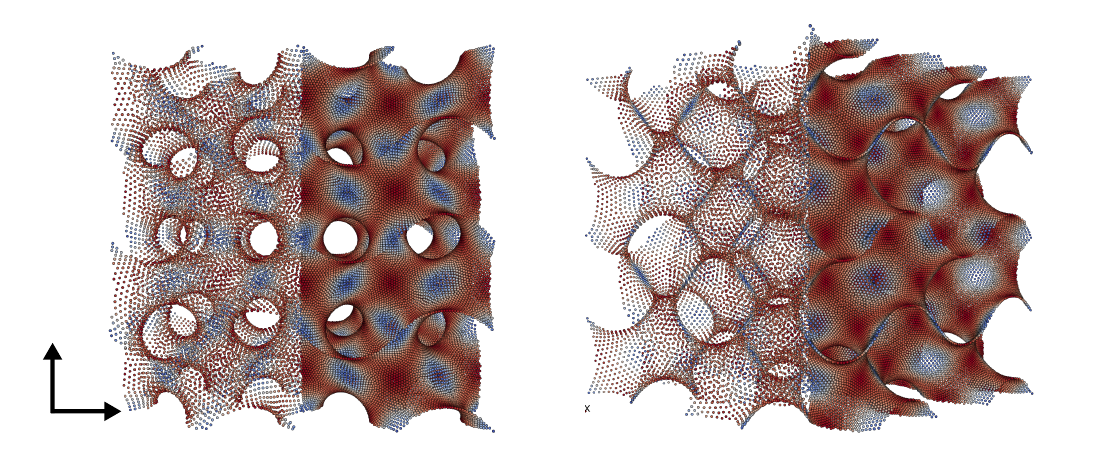
    \caption{Point discretization of the 3D triply periodic minimal surface (TPMS) gyroid over $2\times 2\times 2$ unit cells with periodic boundary conditions and curvature adaptivity $\tau\in\{0.0,\, 0.5\}$. The point distributions for different degrees of curvature adaptivity $\tau$ are shown in two halves of the domain as indicated by the title labels. The two panels show different views of the same result. The points are colored according to the local deviation from flatness of the surface (color bar).}
    \label{fig:tpmsgyroid}
\end{figure}

We test a range of curvature-adaptivity parameters $\tau \in \{0.0,\,  0.1,\, 0.25,\, 0.5\}$.
The resulting discretizations have increasingly many particles, $n_p\in \{27\,214,\, 41\,558,\, 64\,542,\, 108\,178\}$ with a maximum deviation from $h_i$ of $\|e\|_\infty\in \{0.14,\, 0.13,\, 0.27,\, 0.17\}$ and and average deviation of $\|e\|_2\in \{0.023,\, 0.022,\, 0.024,\, 0.041\}$. The two extreme results for this high-genus surface with triply periodic boundary conditions are shown in Fig.~\ref{fig:tpmsgyroid}.

\subsection{Non-parametric curved surfaces}

We demonstrate the applicability of the SAISS algorithm for the adaptive discretization of non-parametric curved surfaces defined by triangulated point clouds. 
To obtain an implicit surface representation as required by SAISS, we use a simple pre-processing step: We use the surface normals associated with the vertices in the data sets to create a narrow-band discretization of the surface by normally extending the vertices into the embedding space by a fixed normal displacement. The signed-distance function values of the narrow-band particles are then initialized with the known distance by which they were shifted in the normal direction. This narrow-band level set is then used as an input to the PCP method, which yields the implicit representation of the surface required by SAISS.

\subsubsection{3D spot data set}

We first consider the \textit{3D spot} data set of a cartoon cow \cite{crane2013robust}. 
The original data set\footnote{available from \url{https://www.cs.cmu.edu/~kmcrane/Projects/ModelRepository}} has a higher vertex density in high-curvature areas, such as the ears and the horns of the cow. It is therefore already curvature-adaptive. To convert the triangulation into an implicit function, we extend all surface vertices in the direction of their normals by four layers into the inside and outside of the closed surface, with a constant spacing of $h_b=0.005$. To account for the varying resolution in the tangent direction, and to ensure sufficient numbers of neighboring particles in near-flat regions, we use a large cutoff radius $r_c=40h_b$ for the PCP method.
This yields smooth curvature fields with deviation from flatness between 0.2 and 55.

We use the 2930 vertices of the triangulation as initial particle set for SAISS 
with $h_0=0.025$ and $\kappa_\text{ref}=0.2$ (the minimum deviation from flatness anywhere on the surface). All other parameters of the SAISS method remain the same as in the 3D Gaussian bump case above.

\begin{figure}
    \centering
    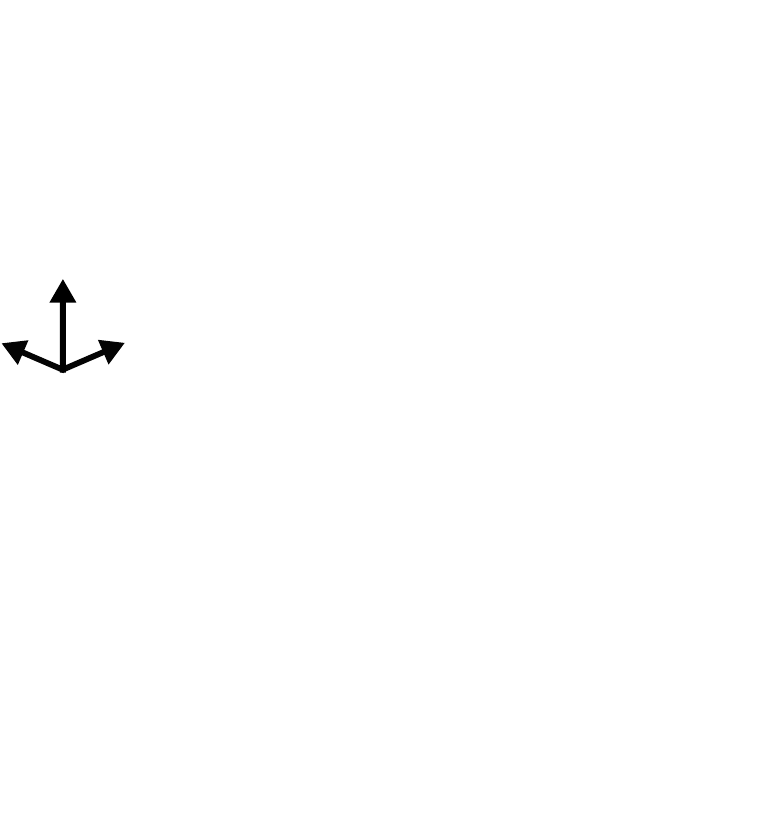
    \caption{Resulting particle distributions on the 3D spot surface for different values of the curvature-adaptivity parameter $\tau$ (panel labels). The scene is clipped at the $y-z$ plane for better visualization. The particles are colored according to the local deviation from flatness of the surface (color bar), as computed by the PCP method.}
    \label{fig:3Dspotdiscretization}
\end{figure}

We compute particle distributions for $\tau\in\{0.0, 0.1, 0.25, 0.5\}$, where $\tau=0$ removes the curvature-adaptivity from the data set. The results are shown in Fig.~\ref{fig:3Dspotdiscretization}. The final discretizations have between 8702 ($\tau=0$) and 35\,517 ($\tau=0.5$) particles. The average relative deviation of the nearest-neighbor distance from the characteristic length ranges from 0.028 to 0.036. 
The bias toward higher particle density is more pronounced for larger $\tau$. For $\tau=0.5$, the smallest neighbor distance is again about a factor of two smaller than required, as in the Gaussian bump case.

\subsubsection{Stanford bunny}

As a second non-parametric surface, we consider the \textit{Stanford bunny}\footnote{available from \url{http://graphics.stanford.edu/data/3Dscanrep}}, which was acquired by 3D scanning of a real-world object. It has very high curvature around the ears and at the bottom, making it a good test case for the robustness of the SAISS method. In its original version, the surface has holes. Since the SAISS method is designed for closed surfaces, we use a closed version of the Stanford bunny\footnote{\url{https://sketchfab.com/3d-models/stanford-bunny-3d-printable-fixed-free-31b33b3a15494df0a13f0a8975d3fa4e}} with 35\,167 vertices, in which sharp angles have been slightly smoothed and holes have been filled. We rescale the coordinate vectors of the vertices by 0.01 to map the surface into the standard domain $[-1,1]^3$. We then transform it to an implicit description using the approach described above with bulk spacing $h_b=0.001$. The surface normals are obtained using the {\tt trimesh} library \cite{trimesh}. For the PCP method, we use a cutoff radius of $r_c=25h_b$. For the SAISS method, we use the same cutoff radii and support thresholds as for the 3D spot data set but choose $h_0=0.01$, $\varepsilon_\text{SAISS}=10^{-8}$, and $\kappa_\text{ref}=0.0$. 

\begin{figure}
    \centering
    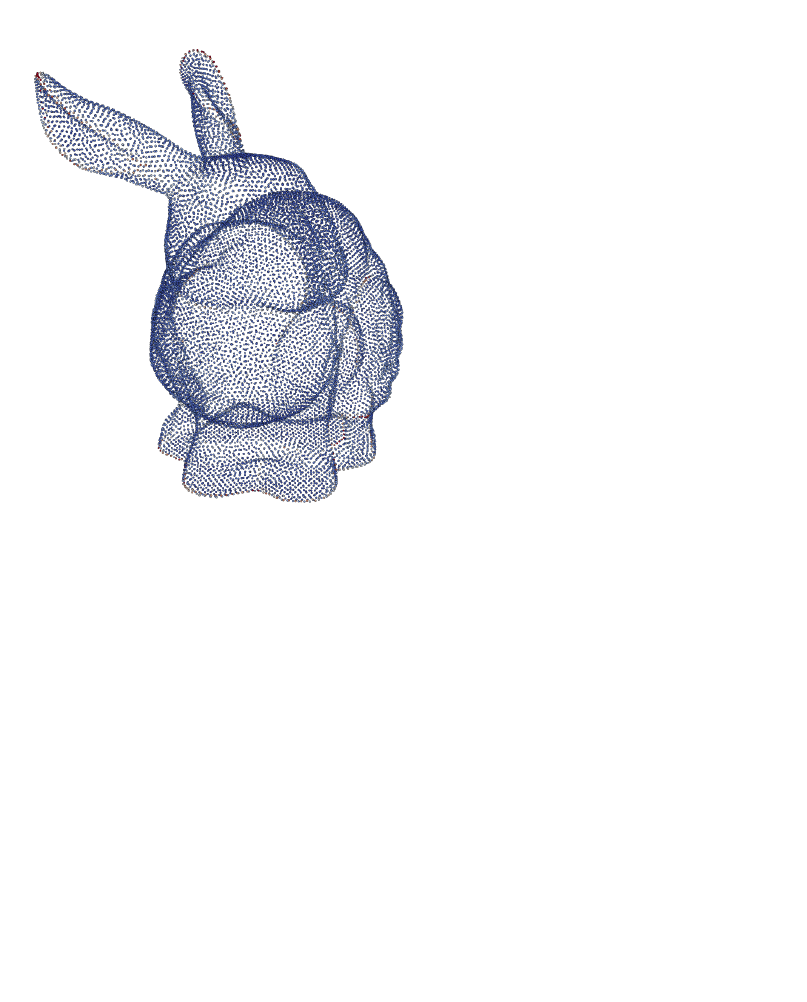
    \caption{Front view of the SAISS particle discretizations of the Stanford bunny for curvature-adaptivity parameters $\tau=0$, $\tau=0.1$, and $\tau=0.25$ (panel labels). The particles are colored according to the local deviation from flatness of the surface. The visualization does not use  textures, yet the features of the bunny, such as its eyes, can be recognized due to the curvature adaptivity.}
    \label{fig:3Dbunnydistribution}
\end{figure}

\begin{figure}
    \centering
\begingroup%
  \makeatletter%
  \providecommand\color[2][]{%
    \errmessage{(Inkscape) Color is used for the text in Inkscape, but the package 'color.sty' is not loaded}%
    \renewcommand\color[2][]{}%
  }%
  \providecommand\transparent[1]{%
    \errmessage{(Inkscape) Transparency is used (non-zero) for the text in Inkscape, but the package 'transparent.sty' is not loaded}%
    \renewcommand\transparent[1]{}%
  }%
  \providecommand\rotatebox[2]{#2}%
  \newcommand*\fsize{\dimexpr\f@size pt\relax}%
  \newcommand*\lineheight[1]{\fontsize{\fsize}{#1\fsize}\selectfont}%
  \ifx\svgwidth\undefined%
    \setlength{\unitlength}{408.93610447bp}%
    \ifx\svgscale\undefined%
      \relax%
    \else%
      \setlength{\unitlength}{\unitlength * \real{\svgscale}}%
    \fi%
  \else%
    \setlength{\unitlength}{\svgwidth}%
  \fi%
  \global\let\svgwidth\undefined%
  \global\let\svgscale\undefined%
  \makeatother%
  \begin{picture}(1,0.68944898)%
    \lineheight{1}%
    \setlength\tabcolsep{0pt}%
    \put(0,0){\includegraphics[width=\unitlength,page=1]{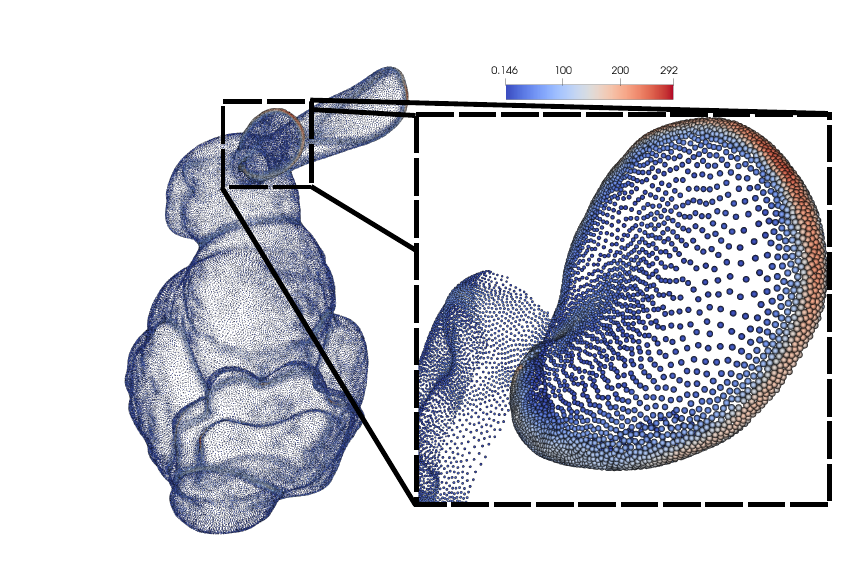}}%
    \put(0.56868126,0.62135015){\color[rgb]{0,0,0}\makebox(0,0)[lt]{\lineheight{1.25}\smash{\begin{tabular}[t]{l}Deviation from Flatness\end{tabular}}}}%
    \put(0,0){\includegraphics[width=\unitlength,page=2]{ear_view_svg-tex.pdf}}%
    \put(0.16456295,0.07352909){\color[rgb]{0,0,0}\transparent{0.76632297}\makebox(0,0)[lt]{\lineheight{1.25}\smash{\begin{tabular}[t]{l}$y$\end{tabular}}}}%
    \put(0.12825969,0.01001333){\color[rgb]{0,0,0}\transparent{0.76632297}\makebox(0,0)[lt]{\lineheight{1.25}\smash{\begin{tabular}[t]{l}$x$\end{tabular}}}}%
    \put(0.09318963,0.11902834){\color[rgb]{0,0,0}\transparent{0.76632297}\makebox(0,0)[lt]{\lineheight{1.25}\smash{\begin{tabular}[t]{l}$z$\end{tabular}}}}%
    \put(0,0){\includegraphics[width=\unitlength,page=3]{ear_view_svg-tex.pdf}}%
  \end{picture}%
\endgroup%

    \caption{Rear view of the particle discretization of the Stanford bunny for curvature-adaptivity parameter $\tau=0.25$, with a closeup of the left ear (inset). The particles are colored according to the local deviation from flatness of the surface. The closeup of the ear is clipped for better visualization of the flat areas, where the particle density is lower.}
    \label{fig:3DbunnyEars}
\end{figure}

\begin{figure}
    \centering
    \def\svgwidth{0.4\textwidth}
\begingroup%
  \makeatletter%
  \providecommand\color[2][]{%
    \errmessage{(Inkscape) Color is used for the text in Inkscape, but the package 'color.sty' is not loaded}%
    \renewcommand\color[2][]{}%
  }%
  \providecommand\transparent[1]{%
    \errmessage{(Inkscape) Transparency is used (non-zero) for the text in Inkscape, but the package 'transparent.sty' is not loaded}%
    \renewcommand\transparent[1]{}%
  }%
  \providecommand\rotatebox[2]{#2}%
  \newcommand*\fsize{\dimexpr\f@size pt\relax}%
  \newcommand*\lineheight[1]{\fontsize{\fsize}{#1\fsize}\selectfont}%
  \ifx\svgwidth\undefined%
    \setlength{\unitlength}{296.34538786bp}%
    \ifx\svgscale\undefined%
      \relax%
    \else%
      \setlength{\unitlength}{\unitlength * \real{\svgscale}}%
    \fi%
  \else%
    \setlength{\unitlength}{\svgwidth}%
  \fi%
  \global\let\svgwidth\undefined%
  \global\let\svgscale\undefined%
  \makeatother%
  \begin{picture}(1,1.53356822)%
    \lineheight{1}%
    \setlength\tabcolsep{0pt}%
    \put(0,0){\includegraphics[width=\unitlength,page=1]{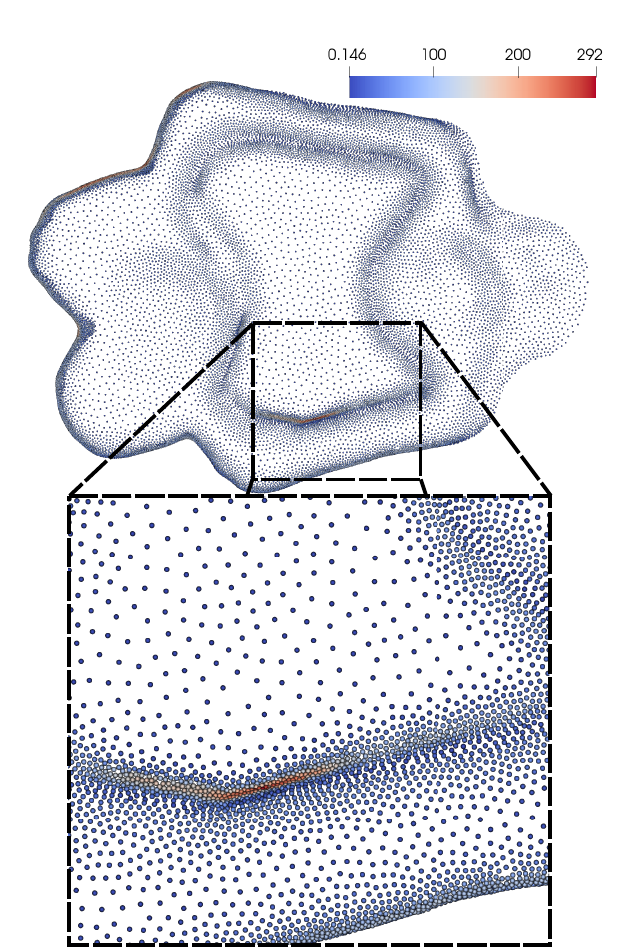}}%
    \put(0.50127,1.47390377){\color[rgb]{0,0,0}\makebox(0,0)[lt]{\lineheight{1.25}\smash{\begin{tabular}[t]{l}Deviation from Flatness\end{tabular}}}}%
    \put(0,0){\includegraphics[width=\unitlength,page=2]{bottom_view_svg-tex.pdf}}%
    \put(0.16087267,1.3462077){\color[rgb]{0,0,0}\makebox(0,0)[lt]{\lineheight{1.25}\smash{\begin{tabular}[t]{l}$x$\end{tabular}}}}%
    \put(0.05372263,1.43311974){\color[rgb]{0,0,0}\makebox(0,0)[lt]{\lineheight{1.25}\smash{\begin{tabular}[t]{l}$y$\end{tabular}}}}%
    \put(0.03470801,1.29691748){\color[rgb]{0,0,0}\makebox(0,0)[lt]{\lineheight{1.25}\smash{\begin{tabular}[t]{l}$z$\end{tabular}}}}%
  \end{picture}%
\endgroup%

    \caption{Bottom view of the particle discretization of the Stanford bunny for curvature-adaptivity parameter $\tau=0.25$. The upper part of the bunny is clipped for better visualization. The particles are colored according to the local deviation from flatness of the surface. The inset shows a closeup of the region of highest (negative) curvature at the border of the bottom indentation.}
    \label{fig:3DbunnyBottom}
\end{figure}

The resulting surface-particle distributions are shown in Fig.~\ref{fig:3Dbunnydistribution} for $\tau=0$, $\tau=0.1$, and $\tau=0.25$. For $\tau=0$, we observe an average deviation from the characteristic length of 0.048 in the final set of $n_p=16\,806$ particles, resulting in a locally regular, down-sampled version of the Stanford bunny. For $\tau=0.1$, we obtain 53\,059 particles with an average resolution error of 0.044. Finally, $\tau=0.25$ yields a roughly three-fold up-sampled version of the surface with 103\,552 particles and an average deviation from the characteristic length of 0.038. Fig.~\ref{fig:3DbunnyEars} and Fig.~\ref{fig:3DbunnyBottom} show closeups to regions of maximal (and rapidly changing) curvature for $\tau=0.25$, confirming that the particle discretization remains smooth, locally regular, and globally adaptive also at the extremal points.

We assess the fidelity with which the original surface is captured by the resulting particle discretizations by computing the asymmetric Hausdorff distance $d_H$, defined as: 
\begin{equation}
    d_H(P,M)=\max_{p\in P}\min_{m\in M}\|p-m\|_2\, ,
\end{equation}
where $M$ is the original surface mesh with nodes $m$, and $P$ is the computed surface point cloud comprising individual points $p$. To compute the unsigned distance $\|p-m\|_2$ between particles and the triangulated surface, we use the raycast from {\tt Open3D} \cite{Zhou2018open3d}. We find Hausdorff distances of $d_H=0.003$ for $\tau=0$, $d_H=0.0036$ for $\tau=0.0036$, and $d_H=0.004$ for $\tau=0.25$. The slight increase with $\tau$ is because the level-set representation is less accurate in high-curvature areas, which are more densely sampled for larger $\tau$. To put the numbers into perspective, these maximum distances correspond to 57\%$\ldots$75\% of the average vertex spacing in the original mesh. The average deviation, however, is only 2.8\%$\ldots$3.98\% of the average vertex spacing for $\tau=0$ and $\tau=0.25$. It is important to keep in mind, though, that the geometric description is piecewise linear in the original mesh, whereas it is a smooth fourth-order polynomial in the geometric representation used by the SAISS method. Some deviations are hence to be expected.

\section{Conclusion and Outlook}

We presented the self-adaptive implicit surface sampling (SAISS) method for computing locally regular and globally adaptive particle discretization of implicitly represented, closed and smooth surfaces according to a prescribed resolution field. Curvature adaptivity is controlled by a user-defined parameter $\tau$. We have shown the applicability of SAISS for adaptively discretizing a variety of parametric and non-parametric surfaces. The SAISS method is an energy-based approach, minimizing a global potential over local particle--particle interactions, accelerated by a line search to find optimal gradient-descent step sizes. SAISS addresses salient issues of energy-relaxation methods by introducing an integral support measure for deciding on particle insertion and removal, and by leveraging the PCP method \cite{schulze2024high} for high-order level-set projections without requiring surface-normal forces

The SAISS method computes high-order approximations of smooth geometric quantities of the curved surface, such as normals and curvatures. These are then used for projecting points and vector-valued quantities into the tangent space, avoiding the need for surface-attractive penalty forces. To determine the number of particles required to cover a given surface with the desired adaptivity, we introduced a particle fusion and addition routine based on estimating the local excess or lack of particles using a continuous integral support measure. This restored convergence of gradient descent but introduced three additional parameters to the method: the support cutoff radius $r_s$ and the lower and upper thresholds $S_\text{min},S_\text{max}$ of the support measure. We found that the standard choice $r_s=2h$ provided good results across a variety of surfaces. The support thresholds, however, required case-specific tuning depending on the spatial curvature gradients. If the curvature changes rapidly compared to the local target length $h_i$, the threshold values should be farther apart to prevent addition--removal loops. We fixed the upper threshold to $S_\text{max}=1.25$ throughout all examples and only changed the lower threshold. We used $S_\text{min}=0.8$ for surfaces with slowly varying curvature and $S_\text{min}=0.7$ if the curvature changes rapidly. For these default parameters, SAISS worked robustly and converged rapidly, often requiring only a few dozen iterations, even if significant down- or up-sampling was necessary. We found the integral support measure to be especially useful when strong curvature adaptivity is required. Then, it outperformed particle insertion and fusion depending on the local number of neighbors and nearest-neighbor distance both in robustness and  quality of the resulting discretization. Despite the stochastic nature of the SAISS algorithm, the final particle set sizes had small coefficient of variation even on the most extreme cases tested.

The stochasticity in the SAISS method helps it escape from local minima, as the optimization problem is highly non-convex with exponentially many (in the number of particles) local minima. For coarse discretization of the unit sphere, the algorithm found the globally optimal arrangement of points. In all other cases, local optima were found for which we quantitatively assessed the quality of the resulting discretizations. In all cases, we started from irregular initial particle distributions far from the optimal solution. The SAISS method reliably converged to high-quality particle distributions that matched the target resolution to within about $\pm$5\%, even for surfaces with wide curvature spectra. The maximum deviations from the characteristic length were found for particles in regions of high curvature gradients, where particles were consistently denser than required. This is by design of the SAISS method and ensures sufficient sampling densities in downstream applications. 

In addition to the bias toward higher particle densities, we also found the SAISS method to sometimes prematurely converge for surfaces with rapidly changing curvature, such as the 3D Gaussian bump. Luckily, premature convergence is easily diagnosed, as it yields particle distributions that are noticeably inhomogeneous. Careful parameter tuning---especially of the solver tolerance $\varepsilon_\text{SAISS}$ and the line search bounds of the gradient descent---was necessary to allow for correct convergence in such cases. 
Premature convergence to bad local minima is well studied in the training of deep neural networks \cite{lecun2002gradient,bottou2010large,smith2017cyclical}. The approaches developed there could potentially also be applied to SAISS, particularly step size schedulers \cite{smith2017cyclical} and stochastic gradient descent \cite{lecun2002gradient,bottou2010large}.

Premature convergence is especially dangerous when the initial particle configuration is far from an optimal configuration. This is, for example, the case when surfaces need to be discretized with high curvature adaptivity starting from an equidistant discretization. It is also the case when computing surface integrals by quadrature, which is highly sensitive to the quality of the point distribution. Quadrature weights can directly be approximated from the per-particle surface areas of the support measure computed by SAISS, as we have shown when computing the total surface area of the sphere. The order of accuracy of this approximation is, however, unclear, and it introduces a smoothing error due to kernel averaging \cite{wendland1995piecewise}. Nevertheless, this might provide a starting point for higher-order quadratures over non-parametric and deforming surfaces.

Ultimately, though, the required quality and regularity of the resulting discretization is dictated by the downstream application. When using SAISS to periodically redistribute surface particles in numerical solvers of PDEs on curved surfaces, premature convergence might not be a problem, as the initial particle distribution (from the previous time step) is already close to optimal \cite{singh2023meshfree,dziuk2013finite}. In this case, we expect that the SAISS method rapidly and robustly yields well-conditioned point distributions. Leveraging the high-order accuracy of the closest-point projections in the PCP method, SAISS can regularize point distributions without distorting the underlying geometry, enabling the adaptive approximation of dynamic curved surfaces by point clouds.

Taken together, the results presented in this paper confirm that the two innovations introduced by SAISS---the integral support measure for deciding on particle insertion and removal, and the high-order level-set projections without normal forces---are able to effectively address two frequent issues of the popular energy-based approach to surface sampling and meshing. Indeed, we found SAISS to converge reliably and rapidly with default parameter settings and to yield numerically well-conditioned surface discretizations with user-controllable curvature adaptivity. This is particularly useful in downstream computational-engineering applications involving surface modeling, analysis, or visualization.

\section*{Declarations}
The authors declare that they have no conflict of interest.

\section*{Acknowledgments}
We thank Philipp H.~Suhrcke and Dr. Alejandra M.~Foggia (both Sbalzarini group) for discussions. We further thank Philipp H.~Suhrcke for providing the starting point for the {\tt OpenFPM} implementation of the SAISS method. This work was supported by the German Federal Ministry of Research, Technology and Space (Bundesministerium f\"{u}r Forschung, Technologie und Raumfahrt; BMFTR) in
the joint project ``6G-life'' (Grant ID 16KISK001K).


\end{document}